\documentclass[bibyear]{aa} 

\usepackage{graphicx}
\usepackage{ amssymb }
\usepackage{multirow}
\usepackage{pdflscape}
\usepackage{natbib}
\bibpunct{(}{)}{;}{a}{}{,} 

\usepackage{txfonts}
\usepackage[
singlelinecheck=false 
]{caption}

\usepackage{lscape} 

\def\f#1   {Fig.~\ref{#1}}
\def\s#1   {Sect.~\ref{#1}}
\def\tab#1   {Table~\ref{#1}}
\def\eq#1   {Eq.~\ref{#1}}
\def\t#1   {Table~\ref{#1}}
\def\comm#1   {{\tt (COMMENT: #1) }}

\def\jup {$J_{\rm up}$}

\begin{document}

 \title{Neutral carbon and highly excited CO in a massive star-forming main sequence galaxy at z=2.2
  }

\author{Drew Brisbin\inst{  1}
\and Manuel Aravena\inst{  1}
\and Emanuele Daddi\inst{2}
\and Helmut Dannerbauer\inst{3,4}
\and Roberto Decarli\inst{5,6}
\and Jorge Gonz\'{a}lez-L\'{o}pez\inst{1}
\and Dominik Riechers\inst{7}
\and Jeff Wagg\inst{8}
}
 
\institute{N\'{u}cleo de Astronom\'{\i}a, Facultad de Ingenier\'{\i}a y Ciencias , Universidad Diego Portales, Av. Ej\'{e}rcito Libertador 441, Santiago, Chile
\and Service d'Astrophysique, CEA Saclay
\and Instituto de Astrof\'{i}sica de Canarias (IAC), E-38205 La Laguna, Tenerife, Spain
\and Universidad de La Laguna, Dpto. Astrof\'{i}sica, E-38205 La Laguna, Tenerife, Spain
\and INAF --- Osservatorio di Astrofisica e Scienza dello Spazio, via Gobetti 93/3. 40129 Bologna, Italy
\and Max-Planck-Institut f\"{u}r Astronomie, K\"{o}nigstuhl, 17, 69117 Heidelberg, Germany
\and Department of Astronomy,  Cornell  University,  220  Space  Sciences Building, Ithaca, NY 14853, USA
\and SKA Organisation, Lower Withington Macclesfield, Cheshire SK11 9DL, UK
}

\authorrunning{ }
\titlerunning{ }

  \abstract{
We used the Plateau De Bure Interferometer to observe multiple CO and neutral carbon transitions in a z=2.2 main sequence disk galaxy, BX610. Our observation of CO(7-6), CO(4-3), and both far-infrared(FIR) [C{\sc i}] lines complements previous observations of H$\alpha$ and low-J CO, and reveals a galaxy that is vigorously forming stars with UV fields (Log($G$ G$_0^{-1}) \lesssim3.25);$     although less vigorously than local ultra-luminous infrared galaxies or most starbursting submillimeter galaxies in the early universe.  Our observations allow new independent estimates of the cold gas mass which indicate $M_\textrm{gas}\sim2\times10^{11}$M$_\odot$, and suggest a modestly larger $\alpha_{\textrm{CO}}$ value of $\sim$8.2.  The corresponding gas depletion timescale is $\sim$1.5 Gyr.  In addition to gas of modest density (Log($n$ cm$^3)\lesssim3$ ) heated by  star formation, BX610 shows evidence for a significant second gas component responsible for the strong high-J CO emission. This second component might either be a high-density molecular gas component heated by star formation in a typical photodissociation region, or could be molecular gas excited by low-velocity C shocks. The  CO(7-6)-to-FIR luminosity ratio we observe is significantly higher than typical star-forming galaxies and suggests that CO(7-6) is not a reliable star-formation tracer in this galaxy.
}


   \maketitle
%

\section{Introduction}\label{sec:intro}
Investigating the interstellar medium (ISM) has long been a crucial part of understanding the processes at work in galaxies and characterizing them. The multitude of easily accessible rotational CO lines have been particularly useful in revealing the density and temperature of molecular gas in galaxies in the local universe. 

Recent advances have made it possible to observe these lines at increasingly early epochs, and there are several samples of galaxies at significant redshift with detections of CO \citep[e.g.,][]{cari2013}. However, currently only a handful of these high-redshift sources have more than a few CO line detections, and those are mostly limited to extreme, high-luminosity systems often undergoing unsustainable starbursts \citep[e.g.,][]{alag2013,cari2013,case2014,both2017,yang2017,ward2018,dann2018}. The existence of a galaxy main sequence has emerged as a useful way of contextualizing galaxy star formation activity. Most star-forming galaxies cluster along a positive trend line in star formation rate (SFR) and stellar mass parameter space with a secondary population of starburst galaxies  with SFRs enhanced by a factor of a few to tens of times larger than their main sequence counterparts with similar stellar masses. The galaxy main sequence persists throughout the local universe and out to high redshifts with an evolving relationship between SFR and stellar mass \citep[e.g.,][]{brin2004,sali2007,elba2007,dadd2007}. Given the abundance of galaxies on the main sequence and the dominance of main sequence galaxies in cosmic luminosity density and star formation rate density during the epoch of peak star formation, many works have examined the relationship between star formation and molecular gas content in main sequence galaxies \citep[e.g.,][]{schm1959,kenn1998,gao2004,krum2009}. 

In this paper we present observations of four emission lines, CO(7-6), CO(4-3), [C{\sc i}] $^3$P$_2$-$^3$P$_1$, and [C{\sc i}] $^3$P$_1$-$^3$P$_0$ (the latter two hereafter denoted [C{\sc i}](2-1) and [C{\sc i}](1-0), respectively), in the massive star-forming galaxy [ESS2003] Q2343-BX610 at z=2.2103, hereafter referred to as BX610. The lines complement previous detections of CO(1-0), and CO(3-2), allowing us to investigate the ISM conditions in this system \citep{tacc2010,arav2014,bola2015}. Unlike most previously detected high-redshift galaxies with an established CO Spectral Line Energy Distribution (SLED), BX610 is thought to be undergoing vigorous star formation (SFR$\sim$60-200 M$_\odot$ yr$^{-1}$) while lying squarely on the z$\sim$2 galaxy main sequence with a stellar mass $M_*\sim10^{11}$M$_\odot$. Few z$\gtrsim$2 main sequence galaxies have been observed in similar gas tracers so far, and even among those few there are likely diverse populations \cite[e.g.,][]{dadd2015,popp2017,tali2018}.  Therefore, understanding the ISM in BX610 is a crucial step in understanding this abundant yet under-investigated galaxy population in the early universe.

In \s{sec:obs} ~ we present our new observations. In \s{sec:results} ~ we present the initial results of our observations and briefly review supporting data from the literature. In \s{sec:disc} ~ we analyze and discuss these observations and their implications for star formation within the galaxy, and in \s{sec:summary} ~ we summarize our findings. Throughout the paper we adopt a flat $\Lambda$CDM cosmology based on Planck 2018 results with $\Omega_\Lambda$=0.6847, $\Omega_\textrm{M}$=0.3153, and $H_0$=67.36 km s$^{-1}$ Mpc$^{-1}$ \citep{plan2018}.

\section{Observations}\label{sec:obs}
Observations of the CO(4-3), [C{\sc i}](1-0), CO(7-6), and [C{\sc i}](2-1) emission lines in BX610 were conducted with the Plateau de Bure Interferometer (PdBI). Observations of the CO(4-3) and [C{\sc i}](1-0) lines were performed on July 11, 15, 21, 26, and 27, 2013, in the 5Dq array configuration (maximum baseline of 97m). Observations of the CO(7-6) and [C{\sc i}](2-1) lines were performed on July 23, 2014, with six available antennas in the D array configuration (maximum baseline of 144m). All observations used the WideX correlator and were pointed toward the position (RA, DECL.)=(23:46:09.43, 12:49:19.21).

Observations of the CO(4-3) and [C{\sc i}](1-0) lines were performed with the 2mm band receivers in two independent setups, tuned to 143.586 and 153.278 GHz, respectively. Observations of the CO(7-6) and [C{\sc i}] 2-1 lines were done with the 1mm receivers, in a single tuning at 251.689 GHz, as both emission lines fall within the 3.6 GHz bandwidth provided by the WideX correlator.

The observing times were 12.5h, 14.3h, and 4.8h for the CO(4-3), [C{\sc i}](1-0), and CO(7-6)/[C{\sc i}](2-1) observations, respectively, for a total of 31.6 hours. All observations were obtained under good weather conditions and phase stability for these frequencies, with typical seeing values between 0.74-1.50'' at 2mm and 0.35'' at 1mm. The bright quasars 2251+158 and 3C454.3 were used for bandpass calibration. The sources 2251+158 and J2327+096 were used for gain calibration at 2mm and 1mm, respectively. The standard calibrator MWC349 and Uranus were used for flux calibration. Extragalactic millimeter calibrator source fluxes can vary significantly over time, and in some cases may be estimated based on fluxes from recent epochs and nearby wavelengths. These methods occasionally result in larger systematic errors than expected, and so calibrator fluxes should be checked carefully. In our examination of the calibration we found a systematic discrepancy between the calculated calibrator fluxes at 150 GHz and the fluxes indicated by the ALMA calibrator archive, with the fluxes used in the PdBI calibration enhanced by 20\%. To correct for this discrepancy we manually corrected our 150 GHz data values by a factor of 0.8. At the observing frequencies, the primary beam FWHMs correspond to 33'' and 20'' at 2mm and 1mm, respectively.

All the data were calibrated using the standard pipeline available in the GILDAS software. The visibilities were inverted and cleaned using a tight box around the initially detected source, down to a threshold of 2.5-sigma. The final cubes yield synthesized beam FWHMs of 3.5''x3.3'' (P.A. 158 deg), 3.3''x3.0'' (P.A. 103 deg), and 2.4''x1.2'' (P.A. 197 deg) with rms noise levels of 0.9, 0.6, and 1.1 mJy beam$^{-1}$ per channel, with channel widths of 50, 50, and 45 km s$^{-1}$ for the CO(4-3), [C{\sc i}] 1-0, and CO(7-6)/[C{\sc i}] 2-1 observations, respectively.
%


\section{Results}\label{sec:results}

\begin{figure*}[htb]
\centering
\includegraphics[height=0.95\textwidth,trim=0.2cm -.2cm -.7cm -.9cm, clip=true,angle=90]{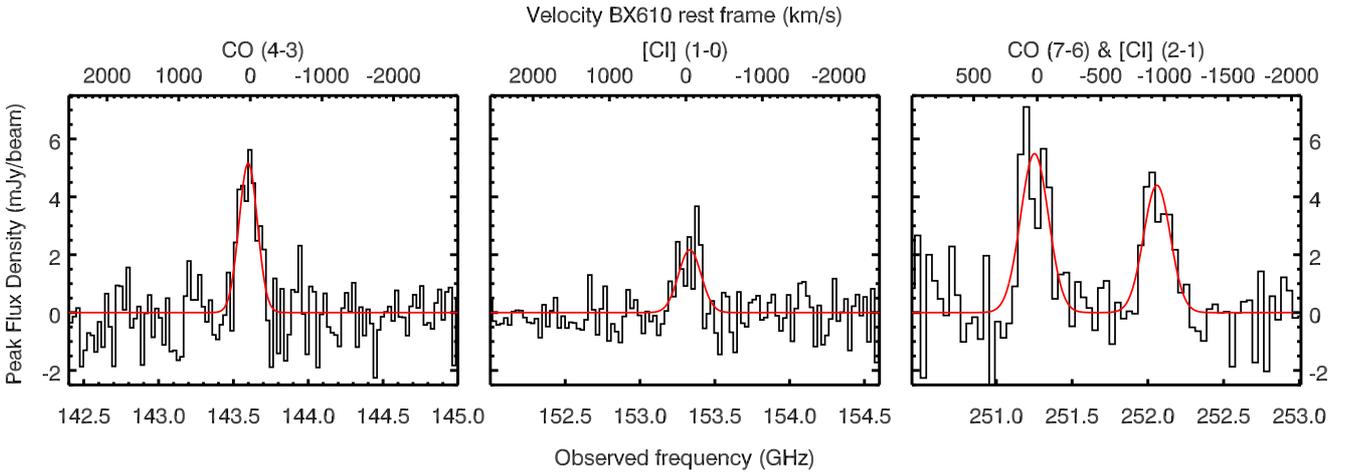}         
        \caption{From left to right, CO(4-3), [C{\sc i}](1-0), CO(7-6) + [C{\sc i}](2-1) peak line spectra in BX610. Bottom axes give observed frequency and top axes give velocity with respect to BX610 rest frame. Respective channel widths for each spectrum are 50, 50, and 45 km s$^{-1}$. Velocities are based on a redshift of z=2.2103 and the velocity in the third panel is based on CO(7-6).  Gaussian fits to each line are shown in red. For CO(7-6) + [C{\sc i}](2-1) the Gaussian parameters were allowed to vary independently and their combined flux densities were fit to the spectrum.
        }
\label{fig:masterspec}
\end{figure*}

We analyzed our data in MAPPING, part of the GILDAS software suite. Based on UV plane source fitting, we determined all line and continuum emission to be consistent with a point source. The continuum flux density at 250 GHz is determined based on the channels below the CO(7-6) frequency and the channels above the [C{\sc i}](2-1) frequency ($\nu \leq$251.009 and $\nu \geq$252.256 GHz). The CO(7-6) and [C{\sc i}](2-1) lines cover spectral ranges 251.085$\leq \nu \leq$251.632 GHz and 251.632$< \nu \leq$252.218 GHz, respectively. Fit results for the continuum and continuum-subtracted emission lines are given in Table \ref{tab:basicobs}. The line spectra of the peak pixel (with continuum subtracted) are shown in Fig. \ref{fig:masterspec}. We also fit the peak line emission with two Gaussians to characterize their velocity distributions. Central frequencies, FWHMs, and amplitudes were allowed to vary independently for both lines. In both cases we find FWHM$\sim$250 km s$^{-1}$ and central frequencies consistent with the expected redshift of z=2.2103.


The continuum emission is significantly weaker at 150 GHz and is difficult to characterize in the CO(4-3) band, and therefore we use only the higher sensitivity data from the [C{\sc i}](1-0) observation to estimate continuum. The continuum estimate is based on channels outside the observed line emission ($\nu \leq$ 152.971 and $\nu \geq$ 153.636 GHz). The line emission shown in Fig. \ref{fig:masterspec} was fit in the same manner as CO(7-6) and [C{\sc i}](2-1). Fitting Gaussians to the peak pixel spectra of [C{\sc i}](1-0) and CO(4-3), we find FWHM of 349 $\pm$ 52 and 301 $\pm$ 33 km s$^{-1}$, respectively.


In addition to the new observations presented here, we also make use of prior CO observations. BX610 was observed and detected in CO(3-2) as part of the Plateau de Bure HIgh-z Blue Sequence Survey (PHIBSS; Tacconi et al. 2010, 2013), and was reanalyzed by \citet{bola2015} as part of a larger sample. The CO(1-0) line was strongly detected by \citet{arav2014} and then again detected by \citet{bola2015} using the Karl G. Jansky Very Large Array. These line estimates are presented with our new observations in Table \ref{tab:basicobs}. We have included an additional 10\% uncertainty on all line and continuum estimates, added in quadrature with the statistical uncertainty, to account for possible flux calibration differences between surveys.

\setlength\tabcolsep{1.5pt}
\begin{table}[htb]
\tiny
\caption{BX610 observed quantities}
\label{tab:basicobs}
\centering
\begin{tabular}{l c c c}
\hline\hline
 Parameter & Unit & Value & Reference \\
\hline
RA      &       hh:mm:ss.ss     & 23:46:09.43   &       \citep{erb2006_1}       \\
Dec.    &       dd:mm:ss.ss             & 12:49:19.21   &       \citep{erb2006_1} \\
redshift &      z                       &       2.2103  &       \citep{fors2009} \\
CO(1-0) & Jy km s$^{-1}$ & 0.105 $\pm$ 0.015 & \citep{bola2015} \\
FWHM & km s$^{-1}$ & 294 $\pm$ 49 & \citep{bola2015} \\
CO(3-2)$^*$ & Jy km s$^{-1}$ & 0.87 $\pm$ 0.11  & \citep{bola2015} \\
FWHM & km s$^{-1}$ & 266 $\pm$ 35 & \citep{bola2015} \\
CO(4-3) & Jy km s$^{-1}$ & 1.82 $\pm$ 0.26 & This work \\
FWHM & km s$^{-1}$ & 301 $\pm$ 33 & This work \\
CO(7-6) & Jy km s$^{-1}$ & 1.88 $\pm$ 0.28 & This work \\
FWHM & km s$^{-1}$ & 258 $\pm$ 44 & This work \\
$[\textrm{C{\sc i}}]$(1-0) & Jy km s$^{-1}$ & 0.80 $\pm$ 0.15  & This work \\
FWHM & km s$^{-1}$ & 349 $\pm$ 52 & This work \\
$[\textrm{C{\sc i}}]$(2-1) & Jy km s$^{-1}$ & 1.34 $\pm$ 0.24 & This work \\
FWHM & km s$^{-1}$ & 247 $\pm$ 54 & This work \\
$S_{150\textrm{GHz}}$   & mJy & 0.41 $\pm$ 0.09  & This work \\
$S_{250\textrm{GHz}}$ & mJy & 2.39 $\pm$ 0.31 & This work \\
\hline
\end{tabular}
\\
\raggedright $^*$Previous value from \citet{tacc2010} is 0.95$\pm$0.08.

\end{table}

\section{Analysis}\label{sec:disc}



\subsection{Infrared luminosity and SFR estimates}
\label{sec:irsfr}
\begin{figure}[htb]
\centering
\includegraphics[height=0.45\textwidth,trim=.7cm .83cm 1.6cm .6cm, clip=true,angle=90]{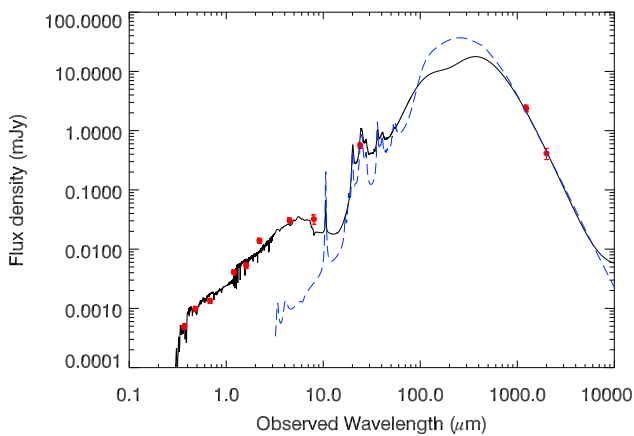}
        \caption{BX610 SED. The MAGPHYS-derived SED is shown as the solid black line. The \citet{drai2007} SED template corresponding to the best fit for massive main sequence 2.0$<z<$2.5 from \citet{beth2015} is shown as the blue dashed line (fit only to the three longest wavelength observations). Observations overlaid as red circles.
        }
\label{fig:sed}
\end{figure}

Previous estimates of SFR in BX610 vary by factors of approximately three but consistently confirm its place on the main sequence. \citet{tacc2013} estimated the SFR of BX610 to be 212$\pm$74 M$_\odot$ yr$^{-1}$, based on combined H$\alpha$ and UV observations.  More recently the SFR has been estimated as 60 M$_\odot$ yr$^{-1}$ based on evolutionary synthesis modeling of the UV-near-infrared(NIR) spectral energy distribution (SED), and 115 M$_\odot$ yr$^{-1}$ based on newly acquired adaptive-optics-assisted H$\alpha$ observations \citep{fors2014,fors2018}.  Although these estimates have all accounted for dust extinction through correction factors or implicit dust models, they should be further compared and tested for consistency with an SED-derived estimate which includes observations of the submillimeter continuum arising from the dust itself. We do a joint UV--submillimeter photometry fit with the high-redshift Multi-wavelength Analysis of Galaxy Physical Properties (MAGPHYS) SED  analysis tool \citep{dacu2008} using our continuum measurements at 150 and 250 GHz, U, G, R, J, and K$_s$ photometry from \citet{erb2006_1,erb2006_2}, HST NIC2 F160W photometry \citep{fors2011_1}, and Spitzer IRAC (4.5 and 8 $\mu$m) and MIPS 24 $\mu$m data extracted from the NASA/ IPAC Infrared Science Archive. The resulting best-fit SED is shown in Fig. \ref{fig:sed}.  The MAGPHYS-derived SFR is Log(SFR (M$_\odot$/yr)$^{-1}$)=2.15$\pm$0.1 and infrared luminosity (8-1000 $\mu$m) is Log(L$_{\textrm{IR}}$ L$_\odot^{-1}$)=12.63$\pm$0.10. We also make use of the 30-1000 $\mu$m luminosity, estimated to be Log(L$_{\textrm{30-1000}}$ L$_\odot^{-1}$)=12.50$\pm$0.10 and the FIR luminosity (42.5-122.5 $\mu$m) Log(L$_{\textrm{FIR}}$ L$_\odot^{-1}$)=12.27$\pm$0.10. In this case the formal errors derived from MAGPHYS are very small due to the discrete nature of the SED templates included in the fitting library. The resulting likelihood distributions of the relevant parameters have a single value dominating the total integrated likelihood power. Therefore, we have taken the sampling spacing of 0.1 in log space as the uncertainty.
~These results are summarized in Table \ref{tab:derivedobs} along with other derived parameters. 

\setlength\tabcolsep{1.5pt}
\begin{table}[htb]
\tiny
\caption{BX610 derived parameters}
\label{tab:derivedobs}
\centering
\begin{tabular}{l c c c}
\hline\hline
 Parameter & Unit & Value & Reference \\
\hline
Log(L$_{\textrm{IR}}$ L$_\odot^{-1}$) & - & 12.63$\pm$0.10 & This work \\ 
Log(L$_{\textrm{FIR}}$ L$_\odot^{-1}$) & - & 12.27$\pm$0.10 & This work \\ 
Log(SFR (M$_\odot$/yr)$^{-1}$)  & -             & 2.15$\pm$0.10 &       This work \\
$M_{\textrm{gas(CO)}}$  &       10$^{11}$M$_\odot$      &       1.1$\pm$0.1$^*$ &       \citep{bola2015} \\
$M_{\textrm{gas([C{\sc i}])}}$  &       10$^{11}$M$_\odot$      &       2.0$\pm$0.4$^*$ &       This work \\
$M_{\textrm{gas(dust)}}$        &       10$^{11}$M$_\odot$      &       2.4$\pm$0.5$^*$ &       This work \\
$M_{*}$ &       10$^{11}$M$_\odot$              &       1.0$\pm$0.3     &  \citep{tacc2013} \\
$\tau_{\textrm{depletion}}$     & Gyr   & 1.5$\pm$0.3   &       This work \\
$T_{\textrm{ex([C{\sc i}])}}$   & K     & 31.8$\pm$6.9  &       This work \\ \hline
\multicolumn{4}{c}{LVG model parameters}\\ \hline
$T$ & K &30$\pm$4 & This work \\
Log($n$ cm$^3$) & - & 4.3$\pm$0.1 & This work \\
Log([CO]/[C]) & - & 0.47$\pm$0.17 & This work \\ \hline
\multicolumn{4}{c}{Two-component PDR model parameters}\\ \hline
Log($G_1$ G$_0^{-1}$) & - & 2.25$\pm$1 & This work \\
Log($n_1$ cm$^3$) & - & 5.0$\pm$0.5 & This work \\ 
Log($G_2$ G$_0^{-1}$) & - & 3.125$\pm$0.375 & This work \\
Log($n_2$ cm$^3$) & - & 3.125$\pm$0.625 & This work \\ 
\hline
\multicolumn{4}{c}{PDR and shock model parameters}\\ \hline
Log($G$ G$_0^{-1}$) & - & 3.25$\pm$0.25 & This work \\
Log($n_{\textrm{PDR}}$ cm$^3$) & - & 3.875$\pm$0.375 & This work \\ 
$v_{\textrm{shock}}$ & km s$^{-1}$ & 10$^\dagger$ & This work \\
Log($n_{\textrm{shock}}$ cm$^3$)& - & 4.3$^\dagger$ & This work \\ 
\hline
\end{tabular}
\\
\raggedright $^*$Error range based on observational uncertainty and neglecting uncertainty in $\alpha_{\textrm{CO}}$, $X[C{\sc I}]/X[H_2]$, and $T_{\textrm{dust}}$.  $^\dagger$Sparsely sampled shock model prevents robust error estimation.
\end{table}

Emission from mid-J CO lines, which usually originates in denser molecular gas, has also been linked to star formation. Several authors have noted the particularly strong correlation between CO(7-6) and infrared luminosity \citep{lu2014,lu2015,liu2015,yang2017}. 
From the relationship presented by (Lu et al. 2015; their equation 1) the CO(7-6) in BX610 would indicate a much larger infrared luminosity, L$_{\textrm{IR}}$=1.2$\times$10$^{13}$L$_\odot$, and a corresponding SFR$\sim$1200 M$_\odot$ yr$^{-1}$ after correcting for a Chabrier initial mass function --- strikingly different from ours and previous SFR estimates. Although the MAGPHYS fit, which uses energy balance considerations to consistently interpret UV through millimeter SEDs, should provide the best estimate, without a probe of the peak of the FIR dust emission we cannot completely rule out the possibility of additional obscured star formation.
~As a simple test we have also investigated the mid-infrared (MIR) through submillimeter SED including Spitzer MIPS and ALMA photometry using the SED template library from \citet{drai2007}.  The full template library allows a breadth of infrared luminosities, including values exceeding the CO(7-6) predicted infrared luminosity, and is not well constrained with only our three photometric points. Based on this same SED template library, \citet{beth2015} determined that massive main sequence galaxies at 2$<z<2.5$ are best characterized by SEDs with a mean ionization parameter of <U>=22.6. Fitting our photometry with the same template indicates a modestly larger infrared luminosity, Log(L$_{\textrm{IR}}$ L$_\odot^{-1}$)=12.89$\pm$0.03, but this still falls short of what would be expected from the CO(7-6) emission. This SED template is also shown in Fig. \ref{fig:sed}.
~Alternatively, the extreme discrepancy between the CO(7-6) emission and  SFR indications from H$\alpha$ and SED fitting might be caused by non-SFR-related gas heating contributing to the strong CO(7-6) emission. From our MAGPHYS-derived FIR luminosity we find the luminosity ratio, Log($\frac{L_{\textrm{CO}(7-6)}}{L_{\textrm{FIR}}}$)=-4.07$\pm$0.12. This is well above all but a few outliers in the local and high-redshift samples presented in the literature.  The relationship from \citet{lu2015} is based on observations of local galaxies, including both main sequence and starburst galaxies. Although there are fewer observations constraining the relationship at high redshifts, the trend persists and, if anything, shifts to even smaller CO(7-6)-to-L$_{\textrm{FIR}}$ ratios, making BX610 even more of an outlier \citep{yang2017}. 
One of the only sources with a comparable or higher ratio and that has been analyzed with multiple molecular lines is NGC 6240, which is believed to be undergoing significant shock heating \citep{meij2013,lu2014}. 
Due to the extreme discrepancy between SFR as estimated by CO(7-6) and other methods, as well as the unusually high CO-to-infrared luminosity ratio and the possibility of shock excitation contributions, we suggest that CO(7-6) is not a reliable SFR diagnostic in this source.

\subsection{Gas mass estimates}
Cold gas mass is the most fundamental property governing star formation in galaxies \citep[e.g.,][]{schm1959,kenn1998,gao2004,krum2009,both2016}. Composed predominantly of H$_2$ with a modest component of helium, cold gas is both the ultimate fuel for star formation and a motivating force by way of gravitationally compelling molecular cloud collapse. Since it is impossible to directly observe H$_2$ from the ground, alternative tracers are used to determine cold gas masses. All of these methods require various assumptions to arrive at the final gas mass value, and these assumptions can come under question in different environments. Our access to several different gas tracers in BX610 provides us with the opportunity to estimate molecular gas masses in multiple ways, and potentially minimize uncertain assumptions.

Molecular CO, which occurs throughout molecular cores, has long been used as an easily observed tracer of molecular hydrogen \citep{solo2005}. Although the abundance of CO relative to H$_2$ varies significantly across different galactic environments (due largely to variations in metallicity and geometry allowing more or less exposure to dissociating radiation), by assuming a conversion factor, $\alpha_{\textrm{CO}}$, the line luminosity of CO(1-0) is routinely converted into star-forming gas mass (often including helium in addition to H$_2$). This gas mass estimation is directly proportional to $\alpha_{\textrm{CO}}$, which is known to vary widely over different environments, from $\sim$1 in local ultra-luminous infrared galaxies (ULIRGs) to $\gtrsim$50 in low-metallicity systems \citep[e.g.,][]{bola2013}. In treating BX610, \citet{bola2015} assume a conversion factor of $\alpha_{\textrm{CO}}=4.36$M$_\odot$(K km s$^{-1}$ pc$^2$)$^{-1}$, which yields a cold gas mass of (1.1$\pm$0.1)$\times10^{11} \textrm{M}_\odot$. 
This conversion factor (which includes the contribution from helium) is similar to Galactic values, often used as an estimate for high-redshift main sequence galaxies \citep{tacc2010,dadd2010,genz2012}.

Rayleigh-Jeans dust continuum has also been shown to be a useful indicator of total gas mass \citep{sant2010,scov2014,scov2016}.  Dust emission in the optically thin Rayleigh-Jeans tail is proportional to the dust mass, and assuming typical values for dust temperature and opacity allows direct calculation of $M_{\textrm{dust}}$ and therefore $M_{\textrm{gas}}$ with an assumed dust-to-gas ratio.   We use the formula presented in \citet{scov2016_e} with our 150 GHz continuum to estimate the gas mass in BX610. We take the recommended values from \citet{scov2016} for the mass-weighted dust temperature, $T$=25 K, and the dust conversion factor $\alpha_{850 \mu \textrm{m}} \equiv L_{\nu 850\mu \textrm{m}} / M_{\textrm{mol}} = (6.7\pm1.7) \times 10^{19} \textrm{erg s}^{-1} \textrm{Hz}^{-1} \textrm{M}_\odot^{-1}$. We find $M_{\textrm{mol}}=(2.4 \pm 0.5)\times10^{11} \textrm{M}_\odot$, which is significantly higher than the estimated cold gas mass from CO(1-0).

Neutral carbon occurs throughout cold molecular gas, and the optically thin [C{\sc i}] transition has been recognized as a useful indicator of molecular gas in galaxies both locally and at high redshift \citep[e.g.,][]{papa2004,both2017,vale2018}.  With our detections of both [C{\sc i}] lines we are able to directly determine the [C{\sc i}] excitation temperature. From Eq. 3 of \citet{walt2011}:
\begin{equation}
T_{\textrm{ex}}=38.8\times \textrm{ln}(2.11\times \frac{L'_{\textrm{[CI]}(1-0)}}{L'_{\textrm{[CI]}(2-1)}})^{-1}
,\end{equation}
where $L'$ indicates the line luminosity in units of K km s$^{-1}$ pc$^{2}$. We find $T_{\textrm{ex}}=31.8\pm6.9$K. From this we can determine the neutral carbon mass. From Eq. 4 of \citet{walt2011}:
\begin{equation}
M_{\textrm{CI}}=5.706\times10^{-4}Q(T_{\textrm{ex}})\frac{1}{3}\textrm{e}^{23.6/T_{\textrm{ex}}}L'_{\textrm{[CI]}(1-0)}[\textrm{M}_\odot]
,\end{equation}
where $Q(T_{\textrm{ex}}$) is the temperature-dependent partition function, $Q(T_{\textrm{ex}}$)=1+3$\textrm{e}^{-T_1/T_{\textrm{ex}}}$+5$\textrm{e}^{-T_2/T_{\textrm{ex}}}$, 
and $T_1$ and $T_2$ are the the [C{\sc I}] transition energy levels above ground state, 23.6 and 62.5 K, respectively. Based on these equations we find $M_{\textrm{CI}}$=(1.4$\pm$0.3)$\times$10$^7$ M$_\odot$. To determine a cold gas mass we need to use a conversion factor, 
$X[CI]/X[H_2]\equiv M_{\textrm{CI}}/(6 M_{\textrm{H}_2})$. This conversion factor may vary with environment, and in particular with the level of metal enrichment. Although there have been fewer studies constraining this ratio across various environments, \citet{vale2018} found a ratio of $X[CI]/X[H_2]$ in the range 1-13$\times10^{-5}$ in their sample of z$\sim$1.2 main sequence galaxies with a weighted mean of 1.55$\times10^{-5}$. 
This value results in an H$_2$ mass of $(1.5\pm0.3)\times10^{11}$M$_\odot$, and a cold gas mass of $(2.0\pm0.4)\times10^{11}$M$_\odot$ (introducing an additional factor of 1.36 to account for helium to be consistent with previous estimates).


Our neutral-carbon- and dust-derived gas masses are consistent with each other, although they are both significantly higher than the CO-derived cold gas mass, suggesting that either $\alpha_{\textrm{CO}}$ is higher in BX610 than the value used here, or $\alpha_{850}$ is lower and $X[CI]/X[H_2]$ is higher than the values used here. Given the wide range over which $\alpha_{\textrm{CO}}$ is known to vary, and the agreement between dust- and [C{\sc i}]-derived masses, we suggest the former case is most likely. If the true cold gas mass of BX610 is given by the weighted mean of the dust- and [C{\sc i}]-derived cold gas masses, $(2.1\pm0.3)\times10^{11}$M$_\odot$, then the corresponding $\alpha_{\textrm{CO}}$ value is 8.2. This is still within the range of demonstrated values in star-forming galaxies. Taking our weighted mean gas mass and our MAGPHYS-based SFR, we find a gas depletion time of 1.5$\pm$0.3 Gyr consistent with local main sequence star-forming galaxies \citep[e.g.,][]{sain2011,sarg2014,mada2014}. 

\subsection{Emission line modeling}
\label{sec:cosled}
\begin{figure}[!htb]
\centering
\includegraphics[height=0.45\textwidth,trim=.8cm 1.3cm 1.4cm 2.25cm, clip=true,angle=90]{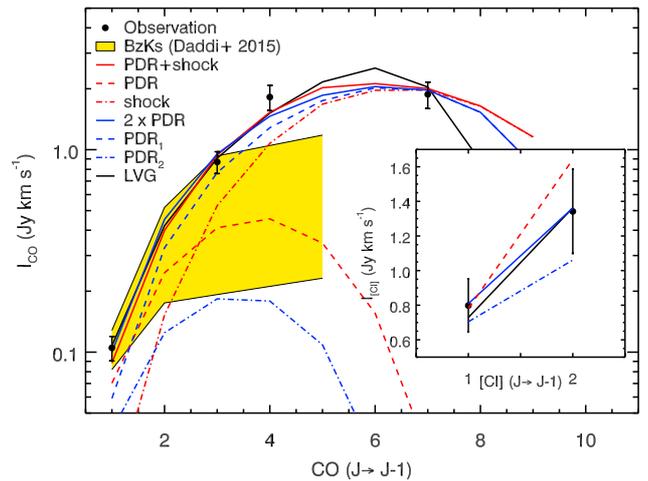}
        \caption{Observations of CO and [C{\sc i}] compared to models. Line observations are given by black circles and corresponding error bars. Yellow swath shows CO SLED for z$\sim$1.5 BzK galaxies from \citet{dadd2015} for comparison (normalized by CO(1-0) or CO(2-1)).  The main figure corresponds to CO observations, while the inset figure corresponds to [C{\sc i}] (note different axis scale). The LVG best-fit model is shown as a black solid line. The PDR plus shock best-fit model is shown as a solid red line with the contributing PDR and shock components shown as dashed and dash-dotted red lines, respectively. The two-PDR-component best-fit model is shown as a solid blue line with the contributing PDR components shown as dashed and dash-dotted blue lines. 
                }
\label{fig:mastersled}
\end{figure}

The integrated CO SLED, comprising our \jup=4 and 7 observations as well as \jup=1 and 3 from \citet{bola2015}, is shown in Fig. \ref{fig:mastersled}.  
Qualitatively we see that the SLED rises strongly through \jup=4 and appears to plateau by \jup=7. Carbon monoxide SLEDs of widely differing shapes have been observed in different sources \citep[e.g.,][]{cari2013,dadd2015}. Although it is not clear that galaxies of a similar class necessarily have similar CO SLEDs, the presence of strong CO emission at \jup$\geq$7 is consistent with intense star formation.  The sample of BzK galaxies at z$\sim$1.5 from \citet{dadd2015} shown in Fig. \ref{fig:mastersled} for example tend to plateau at lower J levels, as do the z$\gtrsim$1 main sequence galaxies in the Hubble Ultra Deep Field detected by \citet{deca2016}. This contrast suggests BX610 has more hot, dense, and highly excited molecular gas, possibly due to enhanced star formation or other mechanisms.

\subsubsection{Large velocity gradient modeling}
As a first attempt to characterize the molecular and atomic gas, we fit the CO SLED and the neutral carbon emission lines with large velocity gradient (LVG) models produced by RADEX \citep{vand2007}. Large velocity gradient models are useful tools for determining bulk gas properties such as density, temperature, and chemical abundances, without specifying any particular power source (such as star formation, AGNs, or shocks for example). While most commonly used to model CO SLEDs, LVG models can also include neutral carbon. Under the assumption that the neutral carbon is tracing the same gas and predominantly excited by molecular hydrogen, as is CO, a single LVG model can simultaneously characterize line emission from both CO and [C{\sc i}] (see e.g, \citet{isra2015}). We use a CO abundance per velocity gradient of 10$^{-5}$ pc (km s$^{-1}$)$^{-1}$, a line width of 300 km s$^{-1}$, and a background temperature of 8.75 K(=2.73 K$\times(1+\textrm{z})$) to produce a grid of CO and [C{\sc i}] emission line models.  Our grid varies the hydrogen density from Log($n$ cm$^3$)=1 to 7 in steps of 0.05, temperature from $T$=10 to 130 in steps of 1 K, and Log([CO]/[C]) abundance ratio from 0.3 to 1.3 in steps of 0.067 (corresponding to [CO]/[C] abundances of 2-20). We compare each emission line model with our observed SLED to find $\chi^2$ values. For the full set of models we convert these $\chi^2$ values to normalized likelihood values ($\mathcal{L}\equiv \frac{\textrm{e}^{-\chi^2/2}}{\sum \textrm{e}^{-\chi^2/2}}$). In Fig. \ref{fig:cornerpdr} we show a corner plot illustrating the likelihood distribution of the LVG parameter space. As seen in the plot, the temperature and density are well constrained with 68\% likelihood ranges of $T$=30$\pm$4 K and Log($n$ cm$^3$)=4.3$\pm$0.1, respectively.  A [CO]/[C] abundance ratio near the low end is preferred in our fit, with a  68\% likelihood range of Log([CO]/[C])=0.47$\pm$0.17. Our best-fit model, with $n$=10$^{4.2}$ cm$^{-3}$, $T$=34 K, and Log([CO]/[C])=0.43 is overlaid on our observations in Fig. \ref{fig:mastersled}. This model provides an adequate fit to the full set of emission lines ($\chi^2_{\rm reduced}=1.04$).

\begin{figure}[!htb]
\centering
\includegraphics[width=0.45\textwidth,trim=-1cm -1cm -.5cm -.5cm, clip=true]{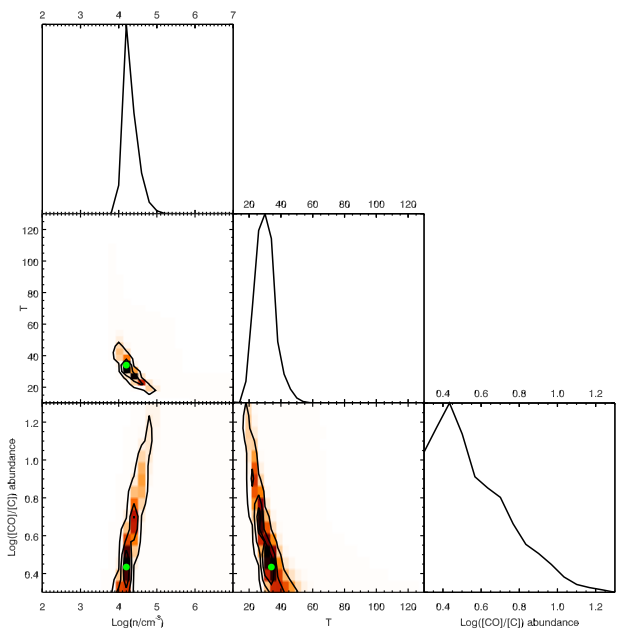}
        \caption{Likelihood corner plots for the parameters in our LVG model. Internal panels give the two-dimensional likelihoods of density, temperature, and [CO]/[C] abundance.  Top right panels give their one-dimensional marginalized likelihoods. Contour levels enclose the 95, 68, and 38\% most likely regions of  each parameter space. A green dot indicates the best-fit model shown in Fig. \ref{fig:mastersled}.
        }
\label{fig:cornerlvg}
\end{figure}

\subsubsection{Photodissociation regions and shocks}
Photodissociation regions (PDRs) powered by the UV emission from young, newly formed stars have long been recognized as an excitation source of infrared line emission and PDR models have proven quite useful in describing individual star formation regions locally, as well as integrated emission from star-formation-dominant galaxies. Furthermore, these models have proven effective at differentiating between normal and starburst galaxies based on their density and UV radiation field intensity \citep[e.g.,][]{malh2001,luhm2003,stac2010,bris2015}. In the case of BX610, PDR models allow us to simultaneously consider both the CO and [C{\sc i}] emission, and also characterize infrared luminosity since much of the UV luminosity from young, newly formed stars is reprocessed by dust into infrared radiation. 

Although multiple different PDR models exist with differing levels of complexity, for our purposes the straightforward semi-infinite slab models of the ``PDR Toolbox'' model suffice  \citep{poun2008,kauf2006}. The parameter space for these models varies gas density from  Log($n$ cm$^3$)=1 to 7 in steps of 0.25, and UV radiation (6 eV$<h\nu<$13.6 eV) field intensity (measured in units of the Habing field, G$_0=1.6\times10^{-3}$ erg cm$^{-2}$ s$^{-1}$) from Log($G$ G$_0^{-1}$)=-0.5 to 6.5 in steps of 0.25. At each density and UV intensity, a line flux is given for each of our observed lines (as well as an infrared continuum flux which varies by UV field but not density), which we then scale up or down uniformly by an effective beam-filling factor to most closely match the absolute values of all our observed emission features. 

\begin{figure}[!htb]
\centering
\includegraphics[width=0.45\textwidth,trim=-1cm -1cm -.5cm -.5cm, clip=true]{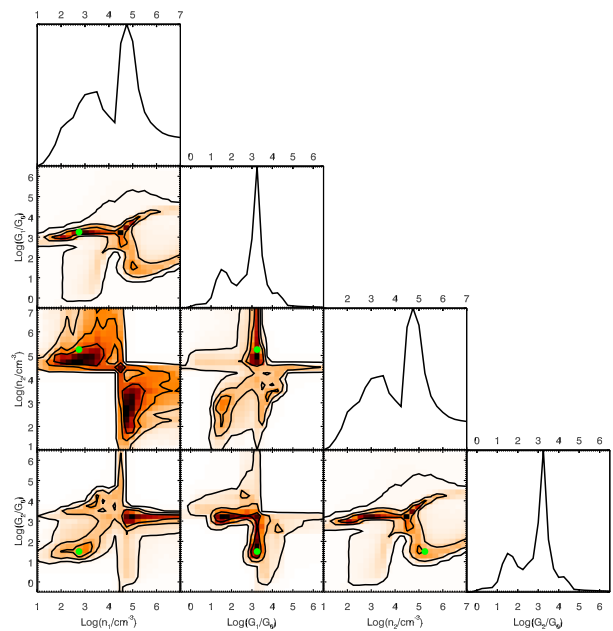}
        \caption{Likelihood corner plots for the parameters in our two-component PDR model. Internal panels give the two-dimensional likelihoods of density, and UV field for each of the two components. Contour levels enclose the 95, 68, and 38\% most likely regions of  each parameter space. The green dot indicates the best-fit model shown in Fig. \ref{fig:mastersled}.
        }
\label{fig:cornerpdr}
\end{figure}

A single component PDR model is not able to adequately fit all four CO lines, both [C{\sc i}] lines, and  infrared luminosity. In general the observed CO lines require relatively higher densities ($n>$10$^4$ cm$^{-3}$) while the [C{\sc i}] lines require a lower-density PDR component ($n<$10$^4$ cm$^{-3}$). 
BX610 may be better represented with two PDR components. This might be the case if BX610 has a phase of higher-density star-forming molecular gas emitting high-J CO emission, and a less-dense but widespread component responsible for much of the [C{\sc i}] and infrared continuum emission. The best two-component PDR model includes a modest density and UV component,  Log($n$ cm$^3$)=3.125$\pm$0.625, Log($G$ G$_0^{-1}$)=3.125$\pm$0.375, which dominates the [C{\sc i}] and infrared luminosity, and a higher-density component, Log($n$ cm$^3$)=5.0$\pm$0.5, Log($G$ G$_0^{-1}$)=2.25$\pm$1.0, which dominates the CO SLED. This two-component model suggests a mixture of gas conditions, but with generally low or modest UV fields, typical of local main sequence galaxies rather than starbursts \citep{malh2001,luhm2003}. This fit is significantly better than a single-component PDR fit, though it is sufficiently far off to suggest that it does not perfectly explain the emission in BX610 ($\chi^2_{\rm reduced}=2.73$). The two-component PDR SLED is plotted in Fig. \ref{fig:mastersled}, and the corner plot of its parameter space is shown in Fig. \ref{fig:cornerpdr}. The predicted infrared luminosity is $L_{30-1000}=3.1\times$10$^{12}$, matching our estimate well.

\begin{figure}[!htb]
\centering
\includegraphics[width=0.45\textwidth,trim=-1cm -1cm -.5cm -.5cm, clip=true]{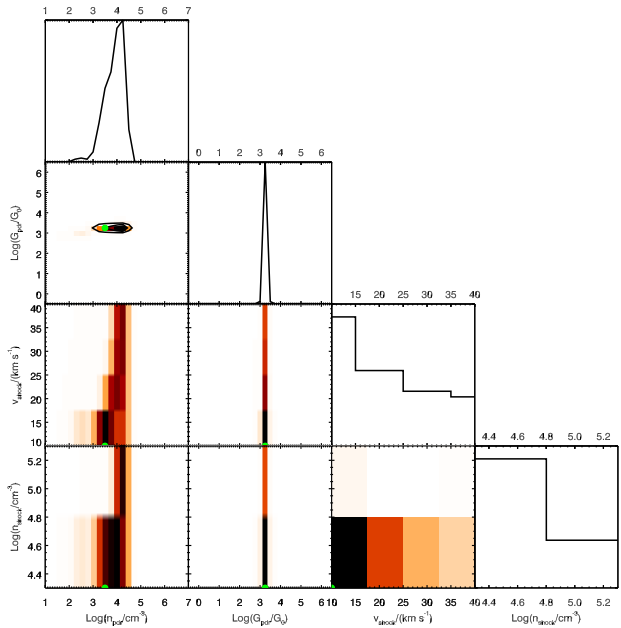}
        \caption{Likelihood corner plots for the parameters in our combined PDR shock model. Internal panels give the two-dimensional likelihoods of PDR density, PDR UV field, shock velocity, and pre-shock density. Where shown, contours enclose the 95, 68, and 38\% most likely regions of parameter space (they have been suppressed for the parameters of the coarsely sampled shock models). The green dot indicates the best-fit model shown in Fig. \ref{fig:mastersled}.
        }
\label{fig:cornershock}
\end{figure}

As noted previously, BX610 has a very high CO(7-6)-to-L$_{\rm FIR}$ ratio, similar to NGC 6240 in which shocks are suspected of powering strong CO emission. Shocks should therefore also be considered as a possible power source for CO emission in BX610. To investigate this possibility we use C shock models from \citet{flow2010}. These models give CO line emission over a small grid of shock velocities ranging from 10 to 40 km s$^{-1}$ in steps of 10 km s$^{-1}$ and at pre-shock densities of 2$\times$10$^4$ and 2$\times$10$^5$ cm$^{-3}$. We use the shock models in conjunction with PDR toolbox models to simultaneously reproduce the CO SLED. 
Dust is not efficiently heated through shocks, so we additionally constrain the PDR model with L$_{\rm FIR}$. Although line emission from neutral carbon could also be enhanced in shocks, the available \citet{flow2010} models do not include [C{\sc i}] emission, so we use the observed [C{\sc i}] lines to constrain the upper limit of PDR-predicted [C{\sc i}] emission. Specifically, a PDR-predicted [C{\sc i}] line at or below the observed flux contributes nothing to the overall $\chi^2$, and a predicted line flux above the observed line contributes in the normal way, weighted by its residual divided by the line flux uncertainty. We emphasize that this is only a suggestive exercise to check the plausibility of shocks contributing significantly to CO emission. Due to the nonlinear nature of this model, as well as the number of free parameters and the dearth of CO lines which constrain shocks, it has no statistical authority. The best fit shown in Fig. \ref{fig:mastersled} matches the observed CO SLED and infrared luminosity well ($L_{\textrm{30-1000}}=3.1\times10^{12}$L$_\odot$), with shocks dominating CO emission at \jup$\geq3$. 
The shock--PDR parameter likelihoods shown as a corner plot in Fig. \ref{fig:cornershock} show well-constrained UV fields, Log($G$ G$_0^{-1}$)=3.25$\pm$0.25, and PDR density, Log($n$ cm$^3$)=3.875$\pm$0.375, which is consistent with our prior LVG analysis. Due to the sparseness of the C shock model grid we cannot robustly estimate shock property uncertainties. Our best-fit model corresponds to a pre-shock density of 2$\times$10$^4$ cm$^{-3}$ and a shock velocity of 10 km s$^{-1}$, though higher velocities and densities cannot be ruled out.

\subsubsection{Physical interpretation}
As BX610 is one of very few examples of z$>2$ main sequence galaxies with detailed gas tracer detections, it is worth investigating the nature of its star formation, and comparing this with the other few examples of this population. Recent work has revealed a population of compact star-forming galaxies (cSFGs) in the galaxy main sequence of the early universe characterized by dense star-forming regions within the cores of massive galaxies \citep[e.g.,][]{nels2014,barr2014,will2014,vand2015,barr2017,tada2017,tali2018,pugl2019}. Although they lie on the main sequence, their star formation densities are more akin to local starbursts, and they are thought to represent a population transitioning from star-forming galaxies to compact quiescent galaxies. 
Compact star-forming galaxies are generally characterized by broad emission lines (FWHM$\sim$600 km s$^{-1}$), compact emission regions ($r_{\textrm{e}}\sim$ a few  kpc), and short depletion times, $\tau_\textrm{dep}\sim100$ Myr. The source 3D-HST GS30274 at z=2.225 has been well studied through PDR-diagnostic lines including several CO and [CI] lines \citep{popp2017}. Analysis by \citet{popp2017} established the PDR parameters of the source as $n$=(6$\pm$0.5)$\times10^4$ cm$^{-3}$ and $G$=(2$\pm$2)$\times10^4G_0$, indicating starburst-like conditions in the compact core. The density and UV field suggested by our PDR+shock model are not particularly extreme; they are consistent with normal star-forming galaxies, though the upper limits of their error ranges do not positively rule out starbursts either \citep{malh2001,luhm2003}. Typical local ULIRGs usually have UV fields Log($G$ G$_0^{-1}$)$\geq$3 and densities Log($n$ cm$^3$)$\geq$4, making the PDR component in our PDR+shock model consistent with both the low-density, low-UV field range of ULIRGs as well as the modestly high-density and UV field range of main sequence galaxies. This leaves open the possibility that BX610 may be forming stars in a scaled up ``normal'' mode of star formation, similar to local non-starburst galaxies, providing an example of  an increasingly important population of galaxies in the universe \citep[see also][]{copp2012,bris2015}. Our two-component PDR model does show that a component with high density, comparable to densities observed in cSFGs, might be present in BX610, though even in this case the second component does not perfectly fit the template of cSFGs since this PDR component also has a relatively low intensity UV field. The emission line widths in BX610, $\sim$300 km s$^{-1}$, are smaller than expected in rapidly rotating cSFGs, and the depletion time is longer as well. 
Furthermore, the effective half light radius for BX610, $r_\textrm{e}$, in the H band was found to be 4.4 kpc \citep{fors2011_2}, significantly exceeding the selection criteria for cSFGs put forward by \citet{vand2015}, which corresponds to a limiting $r_\textrm{e}$ of 2 kpc for BX610.  The combined evidence of the larger spatial extent, lower line velocity widths, longer depletion time, and less extreme PDR parameters in BX610 relative to cSFGs leads us to believe that BX610 is not a cSFG. 

We have used several models to investigate the observed emission in BX610. The LVG model we use yields a robust estimate of the average gas characteristics in the galaxy. It is also the simplest and the most statistically robust model with a $\chi^2_{\rm reduced}$ close to one. Invoking more complicated models is unmerited in that their additional complexity cannot be justified by a corresponding improvement in their fit. This is reflected in the poorer $\chi^2_{\rm reduced}$ of our two-component PDR model, and the shock+PDR model which eludes simple $\chi^2$ analysis. Nonetheless, examining these models may still yield useful physical insight. The simple LVG model for instance says nothing about what mechanism is heating the gas. Photodissociation region models and shock models impose such mechanisms as well as assumptions which further constrain the resulting fits (PDR models for instance assume a stratified temperature structure rather than a single temperature cloud as in the LVG model).

The very high CO(7-6)-to-$L_\textrm{FIR}$ ratio we observe, as well as the consistency of shock models with our CO SLED, suggests shocks may be present in BX610, and raises questions about the nature of these shocks. In local source NGC 6240, with a similarly high CO(7-6)-to-$L_\textrm{FIR}$ ratio, a recent merger event likely caused widespread shock-powered turbulence \citep{meij2013}. A similar event may be the cause in BX610 as well, although there is little evidence to confirm or exclude this possibility. Our unresolved observations provide no morphological information on the dust or molecular gas distribution aside from an upper size constraint, and the emission line profiles are not especially broad, nor are they seen to contain multiple velocity components to indicate potential outflows. \citet{fors2014} however find evidence of a weak or obscured AGN based on MIR colors. They also find broad H$\alpha$ line emission spatially offset from the stellar bulge, which is interpreted as coming from star-formation-powered outflows. This offset could be evidence of a merger interaction in the process of relaxation, though differential dust obscuration could also yield an offset in H$\alpha$ emission.  
Future observations could better disentangle the nature of the shock- and star-formation-excited gas both through improved spatial resolution to look for morphological interaction signatures and by investigating higher-J CO lines and other shocked emission-line diagnostics such as H$_2$O.

\section{Summary}\label{sec:summary}
We used the Plateau de Bure Interferometer to observe CO(4-3), CO(7-6), [C{\sc i}](1-0), and [C{\sc i}](2-1) line emission, as well as the underlying continuum at 150 and 250 GHz in BX610, a star-forming disk galaxy at z=2.2103. Complementing previously detected low-J CO lines, these observations provide insight into a main sequence galaxy in the early universe, which may be representative of a more common population of galaxies than typical starbursting submillimeter galaxies. Our independent measurements of gas mass based on [C{\sc i}] and dust continuum are in good agreement with each other, but are significantly elevated above previous estimates based on CO(1-0), suggesting a modestly higher $\alpha_{\textrm{CO}}$ in BX610 of $\sim$8.2. This is particularly important given the paucity of main sequence galaxies observed in more than one cold gas mass tracer at z$>$1.

The SED of BX610, newly extended with submillimeter photometry, suggests star formation rates of 140$\pm$14 M$_\odot$yr$^{-1}$ , consistent with prior estimates and confirming the existence of this galaxy on the main sequence. Compared to BzK galaxies at z$\sim$1.5, BX610 has an enhanced mid- to high-J CO SLED. One particularly distinguishing feature of BX610 from most local main sequence galaxies is its apparent strong mid-J CO emission compared to $L_{\textrm{FIR}}$: we find Log($\frac{L_{\textrm{CO}(7-6)}}{L_{\textrm{FIR}}}$)=4.06$\pm$0.06. Such a high ratio is known to be shared with only a few other sources, most famously NGC 6240, and is suspected of being a result of shock-enhanced CO emission. Furthermore, a simple single component PDR model is unable to reproduce the observed high-J CO emission without strongly exceeding the observed neutral carbon and infrared luminosity emission. We find that either a second higher-density PDR component or an additional C shock component are needed to fit the emission lines well. Given the empirical correlation between galaxies with  high $\frac{L_{\textrm{CO}(7-6)}}{L_{\textrm{FIR}}}$ ratios and shock excitation, we generally favor the PDR+shock scenario in BX610, although these scenarios  should be investigated further with spatially resolved observations and other shock-diagnostic emission lines. The PDR component of the PDR+shock model is consistent with vigorous star formation occurring in BX610, though not as intense as in off-main sequence starbursts which often show significantly stronger UV fields and shorter gas depletion times. The cause for shock-induced CO emission, if present, is currently uncertain. Whether these shocks are the result of recent interactions or galaxy harassment is uncertain, but should be considered in future investigations.



\section{Acknowledgements}
This work is based in part on observations made with the Spitzer Space Telescope, obtained from the NASA/ IPAC Infrared Science Archive, both of which are operated by the Jet Propulsion Laboratory, California Institute of Technology under a contract with the National Aeronautics and Space Administration.
D.B. acknowledges support from FONDECYT postdoctorado project 3170974.
D.R. acknowledges support from the National Science Foundation under grant number AST-1614213.
H.D. acknowledges financial support from the Spanish Ministry of Economy and Competitiveness (MINECO) under the 2014 Ram\'{o}n y Cajal program MINECO RYC-2014-15686. Este trabajo cont\'o con el apoyo de CONICYT + Programa de Astronom\'ia+ Fondo CHINA-CONICYT.


\bibliographystyle{aa} 
\bibliography{mybib}

\begin{thebibliography}{70}
\expandafter\ifx\csname natexlab\endcsname\relax\def\natexlab#1{#1}\fi

\bibitem[{{Alaghband-Zadeh} {et~al.}(2013){Alaghband-Zadeh}, {Chapman},
  {Swinbank}, {Smail}, {Danielson}, {Decarli}, {Ivison}, {Meijerink}, {Weiss},
  \& {van der Werf}}]{alag2013}
{Alaghband-Zadeh}, S., {Chapman}, S.~C., {Swinbank}, A.~M., {et~al.} 2013,
  \mnras, 435, 1493

\bibitem[{{Aravena} {et~al.}(2014){Aravena}, {Hodge}, {Wagg}, {Carilli},
  {Daddi}, {Dannerbauer}, {Lentati}, {Riechers}, {Sargent}, \&
  {Walter}}]{arav2014}
{Aravena}, M., {Hodge}, J.~A., {Wagg}, J., {et~al.} 2014, \mnras, 442, 558

\bibitem[{{Barro} {et~al.}(2017){Barro}, {Kriek}, {P{\'e}rez-Gonz{\'a}lez},
  {Diaz-Santos}, {Price}, {Rujopakarn}, {Pandya}, {Koo}, {Faber}, {Dekel},
  {Primack}, \& {Kocevski}}]{barr2017}
{Barro}, G., {Kriek}, M., {P{\'e}rez-Gonz{\'a}lez}, P.~G., {et~al.} 2017,
  \apjl, 851, L40

\bibitem[{{Barro} {et~al.}(2014){Barro}, {Trump}, {Koo}, {Dekel}, {Kassin},
  {Kocevski}, {Faber}, {van der Wel}, {Guo}, {P{\'e}rez-Gonz{\'a}lez},
  {Toloba}, {Fang}, {Pacifici}, {Simons}, {Campbell}, {Ceverino},
  {Finkelstein}, {Goodrich}, {Kassis}, {Koekemoer}, {Konidaris}, {Livermore},
  {Lyke}, {Mobasher}, {Nayyeri}, {Peth}, {Primack}, {Rizzi}, {Somerville},
  {Wirth}, \& {Zolotov}}]{barr2014}
{Barro}, G., {Trump}, J.~R., {Koo}, D.~C., {et~al.} 2014, \apj, 795, 145

\bibitem[{{B{\'e}thermin} {et~al.}(2015){B{\'e}thermin}, {Daddi}, {Magdis},
  {Lagos}, {Sargent}, {Albrecht}, {Aussel}, {Bertoldi}, {Buat}, \&
  {Galametz}}]{beth2015}
{B{\'e}thermin}, M., {Daddi}, E., {Magdis}, G., {et~al.} 2015, \aap, 573, A113

\bibitem[{{Bolatto} {et~al.}(2015){Bolatto}, {Warren}, {Leroy}, {Tacconi},
  {Bouch{\'e}}, {F{\"o}rster Schreiber}, {Genzel}, {Cooper}, {Fisher},
  {Combes}, {Garc{\'{\i}}a-Burillo}, {Burkert}, {Bournaud}, {Weiss},
  {Saintonge}, {Wuyts}, \& {Sternberg}}]{bola2015}
{Bolatto}, A.~D., {Warren}, S.~R., {Leroy}, A.~K., {et~al.} 2015, \apj, 809,
  175

\bibitem[{{Bolatto} {et~al.}(2013){Bolatto}, {Wolfire}, \& {Leroy}}]{bola2013}
{Bolatto}, A.~D., {Wolfire}, M., \& {Leroy}, A.~K. 2013, \araa, 51, 207

\bibitem[{{Bothwell} {et~al.}(2017){Bothwell}, {Aguirre}, {Aravena},
  {Bethermin}, {Bisbas}, {Chapman}, {De Breuck}, {Gonzalez}, {Greve},
  {Hezaveh}, {Ma}, {Malkan}, {Marrone}, {Murphy}, {Spilker}, {Strandet},
  {Vieira}, \& {Wei{\ss}}}]{both2017}
{Bothwell}, M.~S., {Aguirre}, J.~E., {Aravena}, M., {et~al.} 2017, \mnras, 466,
  2825

\bibitem[{{Bothwell} {et~al.}(2016){Bothwell}, {Maiolino}, {Cicone}, {Peng}, \&
  {Wagg}}]{both2016}
{Bothwell}, M.~S., {Maiolino}, R., {Cicone}, C., {Peng}, Y., \& {Wagg}, J.
  2016, \aap, 595, A48

\bibitem[{{Brinchmann} {et~al.}(2004){Brinchmann}, {Charlot}, {White},
  {Tremonti}, {Kauffmann}, {Heckman}, \& {Brinkmann}}]{brin2004}
{Brinchmann}, J., {Charlot}, S., {White}, S.~D.~M., {et~al.} 2004, \mnras, 351,
  1151

\bibitem[{{Brisbin} {et~al.}(2015){Brisbin}, {Ferkinhoff}, {Nikola},
  {Parshley}, {Stacey}, {Spoon}, {Hailey-Dunsheath}, \& {Verma}}]{bris2015}
{Brisbin}, D., {Ferkinhoff}, C., {Nikola}, T., {et~al.} 2015, \apj, 799, 13

\bibitem[{{Carilli} \& {Walter}(2013)}]{cari2013}
{Carilli}, C.~L. \& {Walter}, F. 2013, \araa, 51, 105

\bibitem[{{Casey} {et~al.}(2014){Casey}, {Narayanan}, \& {Cooray}}]{case2014}
{Casey}, C.~M., {Narayanan}, D., \& {Cooray}, A. 2014, \physrep, 541, 45

\bibitem[{{Coppin} {et~al.}(2012){Coppin}, {Danielson}, {Geach}, {Hodge},
  {Swinbank}, {Wardlow}, {Bertoldi}, {Biggs}, {Brandt}, {Caselli}, {Chapman},
  {Dannerbauer}, {Dunlop}, {Greve}, {Hamann}, {Ivison}, {Karim}, {Knudsen},
  {Menten}, {Schinnerer}, {Smail}, {Spaans}, {Walter}, {Webb}, \& {van der
  Werf}}]{copp2012}
{Coppin}, K.~E.~K., {Danielson}, A.~L.~R., {Geach}, J.~E., {et~al.} 2012,
  \mnras, 427, 520

\bibitem[{{da Cunha} {et~al.}(2008){da Cunha}, {Charlot}, \&
  {Elbaz}}]{dacu2008}
{da Cunha}, E., {Charlot}, S., \& {Elbaz}, D. 2008, \mnras, 388, 1595

\bibitem[{{Daddi} {et~al.}(2010){Daddi}, {Bournaud}, {Walter}, {Dannerbauer},
  {Carilli}, {Dickinson}, {Elbaz}, {Morrison}, {Riechers}, {Onodera}, {Salmi},
  {Krips}, \& {Stern}}]{dadd2010}
{Daddi}, E., {Bournaud}, F., {Walter}, F., {et~al.} 2010, \apj, 713, 686

\bibitem[{{Daddi} {et~al.}(2015){Daddi}, {Dannerbauer}, {Liu}, {Aravena},
  {Bournaud}, {Walter}, {Riechers}, {Magdis}, {Sargent}, {B{\'e}thermin},
  {Carilli}, {Cibinel}, {Dickinson}, {Elbaz}, {Gao}, {Gobat}, {Hodge}, \&
  {Krips}}]{dadd2015}
{Daddi}, E., {Dannerbauer}, H., {Liu}, D., {et~al.} 2015, \aap, 577, A46

\bibitem[{{Daddi} {et~al.}(2007){Daddi}, {Dickinson}, {Morrison}, {Chary},
  {Cimatti}, {Elbaz}, {Frayer}, {Renzini}, {Pope}, {Alexander}, {Bauer},
  {Giavalisco}, {Huynh}, {Kurk}, \& {Mignoli}}]{dadd2007}
{Daddi}, E., {Dickinson}, M., {Morrison}, G., {et~al.} 2007, \apj, 670, 156

\bibitem[{{Dannerbauer} {et~al.}(2018){Dannerbauer}, {Harrington},
  {Diaz-Sanchez}, {Iglesias-Groth}, {Rebolo}, {Genova-Santos}, \&
  {Krips}}]{dann2018}
{Dannerbauer}, H., {Harrington}, K., {Diaz-Sanchez}, T., {et~al.} 2018, \aj
  [\eprint[arXiv]{1812.03845}]

\bibitem[{{Decarli} {et~al.}(2016){Decarli}, {Walter}, {Aravena}, {Carilli},
  {Bouwens}, {da Cunha}, {Daddi}, {Elbaz}, {Riechers}, {Smail}, {Swinbank},
  {Weiss}, {Bacon}, {Bauer}, {Bell}, {Bertoldi}, {Chapman}, {Colina}, {Cortes},
  {Cox}, {G{\'o}nzalez-L{\'o}pez}, {Inami}, {Ivison}, {Hodge}, {Karim},
  {Magnelli}, {Ota}, {Popping}, {Rix}, {Sargent}, {van der Wel}, \& {van der
  Werf}}]{deca2016}
{Decarli}, R., {Walter}, F., {Aravena}, M., {et~al.} 2016, \apj, 833, 70

\bibitem[{{Draine} \& {Li}(2007)}]{drai2007}
{Draine}, B.~T. \& {Li}, A. 2007, \apj, 657, 810

\bibitem[{{Elbaz} {et~al.}(2007){Elbaz}, {Daddi}, {Le Borgne}, {Dickinson},
  {Alexander}, {Chary}, {Starck}, {Brandt}, {Kitzbichler}, {MacDonald},
  {Nonino}, {Popesso}, {Stern}, \& {Vanzella}}]{elba2007}
{Elbaz}, D., {Daddi}, E., {Le Borgne}, D., {et~al.} 2007, \aap, 468, 33

\bibitem[{{Erb} {et~al.}(2006{\natexlab{a}}){Erb}, {Steidel}, {Shapley},
  {Pettini}, {Reddy}, \& {Adelberger}}]{erb2006_2}
{Erb}, D.~K., {Steidel}, C.~C., {Shapley}, A.~E., {et~al.} 2006{\natexlab{a}},
  \apj, 647, 128

\bibitem[{{Erb} {et~al.}(2006{\natexlab{b}}){Erb}, {Steidel}, {Shapley},
  {Pettini}, {Reddy}, \& {Adelberger}}]{erb2006_1}
{Erb}, D.~K., {Steidel}, C.~C., {Shapley}, A.~E., {et~al.} 2006{\natexlab{b}},
  \apj, 646, 107

\bibitem[{{Flower} \& {Pineau Des For{\^e}ts}(2010)}]{flow2010}
{Flower}, D.~R. \& {Pineau Des For{\^e}ts}, G. 2010, \mnras, 406, 1745

\bibitem[{{F{\"o}rster Schreiber} {et~al.}(2009){F{\"o}rster Schreiber},
  {Genzel}, {Bouch{\'e}}, {Cresci}, {Davies}, {Buschkamp}, {Shapiro},
  {Tacconi}, {Hicks}, {Genel}, {Shapley}, {Erb}, {Steidel}, {Lutz},
  {Eisenhauer}, {Gillessen}, {Sternberg}, {Renzini}, {Cimatti}, {Daddi},
  {Kurk}, {Lilly}, {Kong}, {Lehnert}, {Nesvadba}, {Verma}, {McCracken},
  {Arimoto}, {Mignoli}, \& {Onodera}}]{fors2009}
{F{\"o}rster Schreiber}, N.~M., {Genzel}, R., {Bouch{\'e}}, N., {et~al.} 2009,
  \apj, 706, 1364

\bibitem[{{F{\"o}rster Schreiber} {et~al.}(2014){F{\"o}rster Schreiber},
  {Genzel}, {Newman}, {Kurk}, {Lutz}, {Tacconi}, {Wuyts}, {Bandara}, {Burkert},
  {Buschkamp}, {Carollo}, {Cresci}, {Daddi}, {Davies}, {Eisenhauer}, {Hicks},
  {Lang}, {Lilly}, {Mainieri}, {Mancini}, {Naab}, {Peng}, {Renzini}, {Rosario},
  {Shapiro Griffin}, {Shapley}, {Sternberg}, {Tacchella}, {Vergani},
  {Wisnioski}, {Wuyts}, \& {Zamorani}}]{fors2014}
{F{\"o}rster Schreiber}, N.~M., {Genzel}, R., {Newman}, S.~F., {et~al.} 2014,
  \apj, 787, 38

\bibitem[{{F{\"o}rster Schreiber} {et~al.}(2018){F{\"o}rster Schreiber},
  {Renzini}, {Mancini}, {Genzel}, {Bouch{\'e}}, {Cresci}, {Hicks}, {Lilly},
  {Peng}, {Burkert}, {Carollo}, {Cimatti}, {Daddi}, {Davies}, {Genel}, {Kurk},
  {Lang}, {Lutz}, {Mainieri}, {McCracken}, {Mignoli}, {Naab}, {Oesch},
  {Pozzetti}, {Scodeggio}, {Shapiro Griffin}, {Shapley}, {Sternberg},
  {Tacchella}, {Tacconi}, {Wuyts}, \& {Zamorani}}]{fors2018}
{F{\"o}rster Schreiber}, N.~M., {Renzini}, A., {Mancini}, C., {et~al.} 2018,
  \apjs, 238, 21

\bibitem[{{F{\"o}rster Schreiber} {et~al.}(2011{\natexlab{a}}){F{\"o}rster
  Schreiber}, {Shapley}, {Erb}, {Genzel}, {Steidel}, {Bouch{\'e}}, {Cresci}, \&
  {Davies}}]{fors2011_1}
{F{\"o}rster Schreiber}, N.~M., {Shapley}, A.~E., {Erb}, D.~K., {et~al.}
  2011{\natexlab{a}}, \apj, 731, 65

\bibitem[{{F{\"o}rster Schreiber} {et~al.}(2011{\natexlab{b}}){F{\"o}rster
  Schreiber}, {Shapley}, {Genzel}, {Bouch{\'e}}, {Cresci}, {Davies}, {Erb},
  {Genel}, {Lutz}, {Newman}, {Shapiro}, {Steidel}, {Sternberg}, \&
  {Tacconi}}]{fors2011_2}
{F{\"o}rster Schreiber}, N.~M., {Shapley}, A.~E., {Genzel}, R., {et~al.}
  2011{\natexlab{b}}, \apj, 739, 45

\bibitem[{{Gao} \& {Solomon}(2004)}]{gao2004}
{Gao}, Y. \& {Solomon}, P.~M. 2004, \apj, 606, 271

\bibitem[{{Genzel} {et~al.}(2012){Genzel}, {Tacconi}, {Combes}, {Bolatto},
  {Neri}, {Sternberg}, {Cooper}, {Bouch{\'e}}, {Bournaud}, {Burkert},
  {Comerford}, {Cox}, {Davis}, {F{\"o}rster Schreiber}, {Garcia-Burillo},
  {Gracia-Carpio}, {Lutz}, {Naab}, {Newman}, {Saintonge}, {Shapiro}, {Shapley},
  \& {Weiner}}]{genz2012}
{Genzel}, R., {Tacconi}, L.~J., {Combes}, F., {et~al.} 2012, \apj, 746, 69

\bibitem[{{Israel} {et~al.}(2015){Israel}, {Rosenberg}, \& {van der
  Werf}}]{isra2015}
{Israel}, F.~P., {Rosenberg}, M.~J.~F., \& {van der Werf}, P. 2015, \aap, 578,
  A95

\bibitem[{{Kaufman} {et~al.}(2006){Kaufman}, {Wolfire}, \&
  {Hollenbach}}]{kauf2006}
{Kaufman}, M.~J., {Wolfire}, M.~G., \& {Hollenbach}, D.~J. 2006, \apj, 644, 283

\bibitem[{{Kennicutt}(1998)}]{kenn1998}
{Kennicutt}, Jr., R.~C. 1998, \apj, 498, 541

\bibitem[{{Krumholz} {et~al.}(2009){Krumholz}, {McKee}, \&
  {Tumlinson}}]{krum2009}
{Krumholz}, M.~R., {McKee}, C.~F., \& {Tumlinson}, J. 2009, \apj, 699, 850

\bibitem[{{Liu} {et~al.}(2015){Liu}, {Gao}, {Isaak}, {Daddi}, {Yang}, {Lu}, \&
  {van der Werf}}]{liu2015}
{Liu}, D., {Gao}, Y., {Isaak}, K., {et~al.} 2015, \apjl, 810, L14

\bibitem[{{Lu} {et~al.}(2014){Lu}, {Zhao}, {Xu}, {Gao}, {Armus}, {Mazzarella},
  {Isaak}, {Petric}, {Charmandaris}, {D{\'{\i}}az-Santos}, {Evans}, {Howell},
  {Appleton}, {Inami}, {Iwasawa}, {Leech}, {Lord}, {Sanders}, {Schulz},
  {Surace}, \& {van der Werf}}]{lu2014}
{Lu}, N., {Zhao}, Y., {Xu}, C.~K., {et~al.} 2014, \apjl, 787, L23

\bibitem[{{Lu} {et~al.}(2015){Lu}, {Zhao}, {Xu}, {Gao}, {D{\'{\i}}az-Santos},
  {Charmandaris}, {Inami}, {Howell}, {Liu}, {Armus}, {Mazzarella}, {Privon},
  {Lord}, {Sanders}, {Schulz}, \& {van der Werf}}]{lu2015}
{Lu}, N., {Zhao}, Y., {Xu}, C.~K., {et~al.} 2015, \apjl, 802, L11

\bibitem[{{Luhman} {et~al.}(2003){Luhman}, {Satyapal}, {Fischer}, {Wolfire},
  {Sturm}, {Dudley}, {Lutz}, \& {Genzel}}]{luhm2003}
{Luhman}, M.~L., {Satyapal}, S., {Fischer}, J., {et~al.} 2003, \apj, 594, 758

\bibitem[{{Madau} \& {Dickinson}(2014)}]{mada2014}
{Madau}, P. \& {Dickinson}, M. 2014, Annual Review of Astronomy and
  Astrophysics, 52, 415

\bibitem[{{Malhotra} {et~al.}(2001){Malhotra}, {Kaufman}, {Hollenbach},
  {Helou}, {Rubin}, {Brauher}, {Dale}, {Lu}, {Lord}, {Stacey}, {Contursi},
  {Hunter}, \& {Dinerstein}}]{malh2001}
{Malhotra}, S., {Kaufman}, M.~J., {Hollenbach}, D., {et~al.} 2001, \apj, 561,
  766

\bibitem[{{Meijerink} {et~al.}(2013){Meijerink}, {Kristensen}, {Wei{\ss}}, {van
  der Werf}, {Walter}, {Spaans}, {Loenen}, {Fischer}, {Israel}, {Isaak},
  {Papadopoulos}, {Aalto}, {Armus}, {Charmandaris}, {Dasyra}, {Diaz-Santos},
  {Evans}, {Gao}, {Gonz{\'a}lez-Alfonso}, {G{\"u}sten}, {Henkel}, {Kramer},
  {Lord}, {Mart{\'{\i}}n-Pintado}, {Naylor}, {Sanders}, {Smith}, {Spinoglio},
  {Stacey}, {Veilleux}, \& {Wiedner}}]{meij2013}
{Meijerink}, R., {Kristensen}, L.~E., {Wei{\ss}}, A., {et~al.} 2013, \apjl,
  762, L16

\bibitem[{{Nelson} {et~al.}(2014){Nelson}, {van Dokkum}, {Franx}, {Brammer},
  {Momcheva}, {Schreiber}, {da Cunha}, {Tacconi}, {Bezanson}, {Kirkpatrick},
  {Leja}, {Rix}, {Skelton}, {van der Wel}, {Whitaker}, \& {Wuyts}}]{nels2014}
{Nelson}, E., {van Dokkum}, P., {Franx}, M., {et~al.} 2014, \nat, 513, 394

\bibitem[{{Papadopoulos} {et~al.}(2004){Papadopoulos}, {Thi}, \&
  {Viti}}]{papa2004}
{Papadopoulos}, P.~P., {Thi}, W.-F., \& {Viti}, S. 2004, \mnras, 351, 147

\bibitem[{{Planck Collaboration} {et~al.}(2018){Planck Collaboration},
  {Aghanim}, {Akrami}, {Ashdown}, {Aumont}, {Baccigalupi}, {Ballardini},
  {Banday}, {Barreiro}, {Bartolo}, {Basak}, {Battye}, {Benabed}, {Bernard},
  {Bersanelli}, {Bielewicz}, {Bock}, {Bond}, {Borrill}, {Bouchet}, {Boulanger},
  {Bucher}, {Burigana}, {Butler}, {Calabrese}, {Cardoso}, {Carron},
  {Challinor}, {Chiang}, {Chluba}, {Colombo}, {Combet}, {Contreras}, {Crill},
  {Cuttaia}, {de Bernardis}, {de Zotti}, {Delabrouille}, {Delouis}, {Di
  Valentino}, {Diego}, {Dor{\'e}}, {Douspis}, {Ducout}, {Dupac}, {Dusini},
  {Efstathiou}, {Elsner}, {En{\ss}lin}, {Eriksen}, {Fantaye}, {Farhang},
  {Fergusson}, {Fernandez-Cobos}, {Finelli}, {Forastieri}, {Frailis},
  {Franceschi}, {Frolov}, {Galeotta}, {Galli}, {Ganga}, {G{\'e}nova-Santos},
  {Gerbino}, {Ghosh}, {Gonz{\'a}lez-Nuevo}, {G{\'o}rski}, {Gratton},
  {Gruppuso}, {Gudmundsson}, {Hamann}, {Handley}, {Herranz}, {Hivon}, {Huang},
  {Jaffe}, {Jones}, {Karakci}, {Keih{\"a}nen}, {Keskitalo}, {Kiiveri}, {Kim},
  {Kisner}, {Knox}, {Krachmalnicoff}, {Kunz}, {Kurki-Suonio}, {Lagache},
  {Lamarre}, {Lasenby}, {Lattanzi}, {Lawrence}, {Le Jeune}, {Lemos},
  {Lesgourgues}, {Levrier}, {Lewis}, {Liguori}, {Lilje}, {Lilley}, {Lindholm},
  {L{\'o}pez-Caniego}, {Lubin}, {Ma}, {Mac{\'{\i}}as-P{\'e}rez}, {Maggio},
  {Maino}, {Mandolesi}, {Mangilli}, {Marcos-Caballero}, {Maris}, {Martin},
  {Martinelli}, {Mart{\'{\i}}nez-Gonz{\'a}lez}, {Matarrese}, {Mauri}, {McEwen},
  {Meinhold}, {Melchiorri}, {Mennella}, {Migliaccio}, {Millea}, {Mitra},
  {Miville-Desch{\^e}nes}, {Molinari}, {Montier}, {Morgante}, {Moss}, {Natoli},
  {N{\o}rgaard-Nielsen}, {Pagano}, {Paoletti}, {Partridge}, {Patanchon},
  {Peiris}, {Perrotta}, {Pettorino}, {Piacentini}, {Polastri}, {Polenta},
  {Puget}, {Rachen}, {Reinecke}, {Remazeilles}, {Renzi}, {Rocha}, {Rosset},
  {Roudier}, {Rubi{\~n}o-Mart{\'{\i}}n}, {Ruiz-Granados}, {Salvati}, {Sandri},
  {Savelainen}, {Scott}, {Shellard}, {Sirignano}, {Sirri}, {Spencer},
  {Sunyaev}, {Suur-Uski}, {Tauber}, {Tavagnacco}, {Tenti}, {Toffolatti},
  {Tomasi}, {Trombetti}, {Valenziano}, {Valiviita}, {Van Tent}, {Vibert},
  {Vielva}, {Villa}, {Vittorio}, {Wandelt}, {Wehus}, {White}, {White},
  {Zacchei}, \& {Zonca}}]{plan2018}
{Planck Collaboration}, {Aghanim}, N., {Akrami}, Y., {et~al.} 2018, arXiv
  e-prints [\eprint[arXiv]{1807.06209}]

\bibitem[{{Popping} {et~al.}(2017){Popping}, {Decarli}, {Man}, {Nelson},
  {B{\'e}thermin}, {De Breuck}, {Mainieri}, {van Dokkum}, {Gullberg}, {van
  Kampen}, {Spaans}, \& {Trager}}]{popp2017}
{Popping}, G., {Decarli}, R., {Man}, A.~W.~S., {et~al.} 2017, \aap, 602, A11

\bibitem[{{Pound} \& {Wolfire}(2008)}]{poun2008}
{Pound}, M.~W. \& {Wolfire}, M.~G. 2008, in Astronomical Society of the Pacific
  Conference Series, Vol. 394, Astronomical Data Analysis Software and Systems
  XVII, ed. R.~W. {Argyle}, P.~S. {Bunclark}, \& J.~R. {Lewis}, 654

\bibitem[{{Puglisi} {et~al.}(2019){Puglisi}, {Daddi}, {Liu}, {Bournaud},
  {Silverman}, {Circosta}, {Calabr{\`o}}, {Aravena}, {Cibinel}, {Dannerbauer},
  {Delvecchio}, {Elbaz}, {Gao}, {Gobat}, {Jin}, {Le Floc'h}, {Magdis},
  {Mancini}, {Riechers}, {Rodighiero}, {Sargent}, {Valentino}, \&
  {Zanisi}}]{pugl2019}
{Puglisi}, A., {Daddi}, E., {Liu}, D., {et~al.} 2019, arXiv e-prints
  [\eprint[arXiv]{1905.02958}]

\bibitem[{{Saintonge} {et~al.}(2011){Saintonge}, {Kauffmann}, {Wang}, {Kramer},
  {Tacconi}, {Buchbender}, {Catinella}, {Graci{\'a}-Carpio}, {Cortese},
  {Fabello}, {Fu}, {Genzel}, {Giovanelli}, {Guo}, {Haynes}, {Heckman},
  {Krumholz}, {Lemonias}, {Li}, {Moran}, {Rodriguez-Fernandez}, {Schiminovich},
  {Schuster}, \& {Sievers}}]{sain2011}
{Saintonge}, A., {Kauffmann}, G., {Wang}, J., {et~al.} 2011, \mnras, 415, 61

\bibitem[{{Salim} {et~al.}(2007){Salim}, {Rich}, {Charlot}, {Brinchmann},
  {Johnson}, {Schiminovich}, {Seibert}, {Mallery}, {Heckman}, {Forster},
  {Friedman}, {Martin}, {Morrissey}, {Neff}, {Small}, {Wyder}, {Bianchi},
  {Donas}, {Lee}, {Madore}, {Milliard}, {Szalay}, {Welsh}, \& {Yi}}]{sali2007}
{Salim}, S., {Rich}, R.~M., {Charlot}, S., {et~al.} 2007, \apjs, 173, 267

\bibitem[{{Santini} {et~al.}(2010){Santini}, {Maiolino}, {Magnelli}, {Silva},
  {Grazian}, {Altieri}, {Andreani}, {Aussel}, {Berta}, {Bongiovanni},
  {Brisbin}, {Calura}, {Cava}, {Cepa}, {Cimatti}, {Daddi}, {Dannerbauer},
  {Dominguez-Sanchez}, {Elbaz}, {Fontana}, {F{\"o}rster Schreiber}, {Genzel},
  {Granato}, {Gruppioni}, {Lutz}, {Magdis}, {Magliocchetti}, {Matteucci},
  {Nordon}, {P{\'e}rez Garcia}, {Poglitsch}, {Popesso}, {Pozzi}, {Riguccini},
  {Rodighiero}, {Saintonge}, {Sanchez-Portal}, {Shao}, {Sturm}, {Tacconi}, \&
  {Valtchanov}}]{sant2010}
{Santini}, P., {Maiolino}, R., {Magnelli}, B., {et~al.} 2010, \aap, 518, L154

\bibitem[{{Sargent} {et~al.}(2014){Sargent}, {Daddi}, {B{\'e}thermin},
  {Aussel}, {Magdis}, {Hwang}, {Juneau}, {Elbaz}, \& {da Cunha}}]{sarg2014}
{Sargent}, M.~T., {Daddi}, E., {B{\'e}thermin}, M., {et~al.} 2014, \apj, 793,
  19

\bibitem[{{Schmidt}(1959)}]{schm1959}
{Schmidt}, M. 1959, \apj, 129, 243

\bibitem[{{Scoville} {et~al.}(2014){Scoville}, {Aussel}, {Sheth}, {Scott},
  {Sanders}, {Ivison}, {Pope}, {Capak}, {Vanden Bout}, {Manohar}, {Kartaltepe},
  {Robertson}, \& {Lilly}}]{scov2014}
{Scoville}, N., {Aussel}, H., {Sheth}, K., {et~al.} 2014, \apj, 783, 84

\bibitem[{{Scoville} {et~al.}(2016{\natexlab{a}}){Scoville}, {Sheth}, {Aussel},
  {Vanden Bout}, {Capak}, {Bongiorno}, {Casey}, {Murchikova}, {Koda},
  {{\'A}lvarez-M{\'a}rquez}, {Lee}, {Laigle}, {McCracken}, {Ilbert}, {Pope},
  {Sanders}, {Chu}, {Toft}, {Ivison}, \& {Manohar}}]{scov2016}
{Scoville}, N., {Sheth}, K., {Aussel}, H., {et~al.} 2016{\natexlab{a}}, \apj,
  820, 83

\bibitem[{{Scoville} {et~al.}(2016{\natexlab{b}}){Scoville}, {Sheth}, {Aussel},
  {Vanden Bout}, {Capak}, {Bongiorno}, {Casey}, {Murchikova}, {Koda},
  {{\'A}lvarez-M{\'a}rquez}, {Lee}, {Laigle}, {McCracken}, {Ilbert}, {Pope},
  {Sanders}, {Chu}, {Toft}, {Ivison}, \& {Manohar}}]{scov2016_e}
{Scoville}, N., {Sheth}, K., {Aussel}, H., {et~al.} 2016{\natexlab{b}}, \apj,
  824, 63

\bibitem[{{Solomon} \& {Vanden Bout}(2005)}]{solo2005}
{Solomon}, P.~M. \& {Vanden Bout}, P.~A. 2005, \araa, 43, 677

\bibitem[{{Stacey} {et~al.}(2010){Stacey}, {Hailey-Dunsheath}, {Ferkinhoff},
  {Nikola}, {Parshley}, {Benford}, {Staguhn}, \& {Fiolet}}]{stac2010}
{Stacey}, G.~J., {Hailey-Dunsheath}, S., {Ferkinhoff}, C., {et~al.} 2010, \apj,
  724, 957

\bibitem[{{Tacconi} {et~al.}(2010){Tacconi}, {Genzel}, {Neri}, {Cox}, {Cooper},
  {Shapiro}, {Bolatto}, {Bouch{\'e}}, {Bournaud}, {Burkert}, {Combes},
  {Comerford}, {Davis}, {Schreiber}, {Garcia-Burillo}, {Gracia-Carpio}, {Lutz},
  {Naab}, {Omont}, {Shapley}, {Sternberg}, \& {Weiner}}]{tacc2010}
{Tacconi}, L.~J., {Genzel}, R., {Neri}, R., {et~al.} 2010, \nat, 463, 781

\bibitem[{{Tacconi} {et~al.}(2013){Tacconi}, {Neri}, {Genzel}, {Combes},
  {Bolatto}, {Cooper}, {Wuyts}, {Bournaud}, {Burkert}, {Comerford}, {Cox},
  {Davis}, {F{\"o}rster Schreiber}, {Garc{\'{\i}}a-Burillo}, {Gracia-Carpio},
  {Lutz}, {Naab}, {Newman}, {Omont}, {Saintonge}, {Shapiro Griffin}, {Shapley},
  {Sternberg}, \& {Weiner}}]{tacc2013}
{Tacconi}, L.~J., {Neri}, R., {Genzel}, R., {et~al.} 2013, \apj, 768, 74

\bibitem[{{Tadaki} {et~al.}(2017){Tadaki}, {Kodama}, {Nelson}, {Belli},
  {F{\"o}rster Schreiber}, {Genzel}, {Hayashi}, {Herrera-Camus}, {Koyama},
  {Lang}, {Lutz}, {Shimakawa}, {Tacconi}, {{\"U}bler}, {Wisnioski}, {Wuyts},
  {Hatsukade}, {Lippa}, {Nakanishi}, {Ikarashi}, {Kohno}, {Suzuki}, {Tamura},
  \& {Tanaka}}]{tada2017}
{Tadaki}, K.-i., {Kodama}, T., {Nelson}, E.~J., {et~al.} 2017, \apjl, 841, L25

\bibitem[{{Talia} {et~al.}(2018){Talia}, {Pozzi}, {Vallini}, {Cimatti},
  {Cassata}, {Fraternali}, {Brusa}, {Daddi}, {Delvecchio}, {Ibar}, {Liuzzo},
  {Vignali}, {Massardi}, {Zamorani}, {Gruppioni}, {Renzini}, {Mignoli},
  {Pozzetti}, \& {Rodighiero}}]{tali2018}
{Talia}, M., {Pozzi}, F., {Vallini}, L., {et~al.} 2018, \mnras, 476, 3956

\bibitem[{{Valentino} {et~al.}(2018){Valentino}, {Magdis}, {Daddi}, {Liu},
  {Aravena}, {Bournaud}, {Cibinel}, {Cormier}, {Dickinson}, {Gao}, {Jin},
  {Juneau}, {Kartaltepe}, {Lee}, {Madden}, {Puglisi}, {Sanders}, \&
  {Silverman}}]{vale2018}
{Valentino}, F., {Magdis}, G.~E., {Daddi}, E., {et~al.} 2018, \apj, 869, 27

\bibitem[{{van der Tak} {et~al.}(2007){van der Tak}, {Black}, {Sch{\"o}ier},
  {Jansen}, \& {van Dishoeck}}]{vand2007}
{van der Tak}, F.~F.~S., {Black}, J.~H., {Sch{\"o}ier}, F.~L., {Jansen}, D.~J.,
  \& {van Dishoeck}, E.~F. 2007, \aap, 468, 627

\bibitem[{{van Dokkum} {et~al.}(2015){van Dokkum}, {Nelson}, {Franx}, {Oesch},
  {Momcheva}, {Brammer}, {F{\"o}rster Schreiber}, {Skelton}, {Whitaker}, {van
  der Wel}, {Bezanson}, {Fumagalli}, {Illingworth}, {Kriek}, {Leja}, \&
  {Wuyts}}]{vand2015}
{van Dokkum}, P.~G., {Nelson}, E.~J., {Franx}, M., {et~al.} 2015, \apj, 813, 23

\bibitem[{{Walter} {et~al.}(2011){Walter}, {Wei{\ss}}, {Downes}, {Decarli}, \&
  {Henkel}}]{walt2011}
{Walter}, F., {Wei{\ss}}, A., {Downes}, D., {Decarli}, R., \& {Henkel}, C.
  2011, \apj, 730, 18

\bibitem[{{Wardlow} {et~al.}(2018){Wardlow}, {Simpson}, {Smail}, {Swinbank},
  {Blain}, {Brandt}, {Chapman}, {Chen}, {Cooke}, {Dannerbauer}, {Gullberg},
  {Hodge}, {Ivison}, {Knudsen}, {Scott}, {Thomson}, {Wei{\ss}}, \& {van der
  Werf}}]{ward2018}
{Wardlow}, J.~L., {Simpson}, J.~M., {Smail}, I., {et~al.} 2018, \mnras, 479,
  3879

\bibitem[{{Williams} {et~al.}(2014){Williams}, {Giavalisco}, {Cassata},
  {Tundo}, {Wiklind}, {Guo}, {Lee}, {Barro}, {Wuyts}, {Bell}, {Conselice},
  {Dekel}, {Faber}, {Ferguson}, {Grogin}, {Hathi}, {Huang}, {Kocevski},
  {Koekemoer}, {Koo}, {Ravindranath}, \& {Salimbeni}}]{will2014}
{Williams}, C.~C., {Giavalisco}, M., {Cassata}, P., {et~al.} 2014, \apj, 780, 1

\bibitem[{{Yang} {et~al.}(2017){Yang}, {Omont}, {Beelen}, {Gao}, {van der
  Werf}, {Gavazzi}, {Zhang}, {Ivison}, {Lehnert}, {Liu}, {Oteo},
  {Gonz{\'a}lez-Alfonso}, {Dannerbauer}, {Cox}, {Krips}, {Neri}, {Riechers},
  {Baker}, {Micha{\l}owski}, {Cooray}, \& {Smail}}]{yang2017}
{Yang}, C., {Omont}, A., {Beelen}, A., {et~al.} 2017, \aap, 608, A144

\end{thebibliography}

\end{document}